# Examining Web Application by Clumping and Orienting User Session Data

Mr.T.Deenadayalan[1], Dr.V.Kavitha[2], Mrs.S.Rajarajeswari[3]

[1]University Department, PG Student, Anna University Tirunelveli[1]
Tirunelveli, Tamilnadu, India

[2] University Department, Asst Professor, Anna University Tirunelveli[2]
Tirunelveli, Tamilnadu, India

[3] University Department, Lecturer, Anna University Tirunelveli[3]
Tirunelveli, Tamilnadu, India

*Abstract:*

*The increasing demand for reliable Web applications gives a central role to Web testing. Most of the existing works are focused on the definition of novel testing techniques, specifically tailored to the Web. However, no attempt was carried out so far to understand the specific nature of Web faults. This paper presents a user session based testing technique that clusters user sessions based on the service profile and selects a set of representative user sessions from each cluster and tailored by augmentation with additional requests to cover the dependence relationships between web pages. The created suite not only can significantly reduce the size of the collected user sessions, also viable to exercise fault sensitive paths. The results demonstrate that our approach consistently detected the majority of known faults using a relatively small number of test cases and will be a powerful system when more and more user sessions are being clustered.*

## 1. Introduction

Web applications have been dramatically adopted in a wide range of software application domains, and many of our daily activities rely on the services provided by them. As a consequence, the qualities of these applications have a great impact on our daily lives. Effective testing of web applications is crucial to provide reliable services for the fast growing demand. Furthermore, web applications often undertake frequent updates and upgrades while in production. Testing must be performed efficiently and be completed within a limited time to avoid service disruptions. Recent studies suggest the use of user session data to create test cases [9] [15]. Elbaum et al. proposed five approaches to apply user sessions in testing. The results of their empirical studies show that in terms of the fault detection capability, the test cases created from user sessions were as good as those created by using white box approaches [9].User-session based testing makes use of field data to create test cases, which has the great potential to efficiently generate test cases that can effectively detect residual faults. However, this approach is relatively new compared to traditional well developed techniques. There are several issues that must be addressed before it can serve as a sole testing method in practice.
(1) For an application that has been in production for a long time, the number of user sessions can be extremely large. Using all of the collected user session data requires much effort to determine which portion of the data can serve as the best representative of the system behaviour. Nevertheless, a vast number of user sessions may not necessarily guarantee good coverage of the expected system behaviour. Applying the Pareto principle in this context, it would not be surprising to observe that 80% of the usages tend to use only 20% of the system services.
(2) On the contrary, a newly deployed application would not have many user session data. As suggested by Elbaum et al. [9] the more user sessions used, the better fault detection rate would be. A mechanism to determine the adequacy of the collected user sessions needs to be developed. (3) Web applications are often continuously evolving, and the previously collected user sessions may no longer be valid for the evolved system. There is a need to analyse the changes and evaluate the validity of the collected user session data. To tackle the problem of overwhelmingly large sized user sessions, Elbaum et al. [9] applied two reduction approaches to select a subset of the user sessions, one based on the function, page-to-page transition, and block coverage criteria, and the other based on the similarity of coverage patterns. Sam path et al. [15] applied a formal concept analysis to cluster user sessions and proposed a set of heuristics for selecting a subset of user sessions based on a concept lattice. Both studies showed that their reduction techniques significantly reduced the size of the user-session data, while missing only a small portion of the seeded faults detected by using the original data. These studies focus mainly on the reduction of the user sessions, while the ability of fault detection using the reduced set is subject to that of the original set of user sessions. Hence, the quality of the collected user sessions is the key for effective user-session based testing. In this paper, we present a methodology that takes into account the quality of the user sessions.

We first use a service profile to partition the user sessions, which can be derived from the functional requirements or user manual, or reverse engineered from the source code. With this approach, the services not requested by any collected user sessions and the invalidated user sessions caused by the change of services can be easily identified. Next, we compute the link and data dependence relationships among the web pages of the application. In each service equivalence class, the smallest set of user sessions that contain all the web pages in the equivalent class will be selected and augmented with additional requests to include their dependent web pages. Finally, if any of the services are not covered by any user sessions, then additional test cases will be created to ensure that all the services will be exercised at least once.

Our approach creates test suites that leverage the field data in the user sessions, yet does not heavily rely on the quality of the collected user sessions to achieve adequacy for effective testing. It ensures that
(1) All the services will be tested;
(2) Dependence relationships between web pages that can potentially introduce faults will be covered by the test cases; and
(3) The number of test cases will be significantly reduced from that of the original user sessions. Most importantly, the created test suite will carry crucial information and be vital for revealing faults. The remainder of the paper is organized as follows:

Section 2 gives an overview of existing work in web testing. In Section 3 we present our problem formulation and approach. Section 4 shows the design of the system and we give our conclusions in Section 5.

## 2. Background

A number of model-based testing approaches have been proposed for web applications. Ricca and Tonella [16] proposed a UML-based navigation model to represent the structures of web applications. Based on the model, a path expression can be generated to facilitate test case generation. A similar model proposed by DiLucca et al. [7] includes unit testing that uses web pages as test units and a function level testing that employs use case diagrams to aid test case generation. Both approaches use reverse engineering to extract UML models from the source code. Andrews et al. [1] proposed an approach based on a hierarchical finite state machine to generate test sequences. To reduce the human effort in test input generation, Elbaum et al. proposed a strategy that makes use of user-session data to test web applications [9].

Three different ways of applying user session data to create test cases were proposed. One is making a direct transformation from a user session to a test case, combining different user sessions to form a test case, and reusing user sessions with form modifications.

Two hybrid methods were suggested to increase fault detection rates: one partially satisfies test requirements with user session data and the other satisfies testing requirements with user-session data and test inputs. Although the hybrid methods attempt to address the adequacy issue of the user-session data, the results of their empirical studies, which used the coverage of functions and basic blocks as the test requirements, did not show any improvement in fault detection rates. In addition, they adopted the test suite reduction technique proposed by Harold et al.[10] to reduce the size of the user sessions by selecting the representative sets of user sessions that cover all functions, page-to-page transitions, and blocks.

A clustering approach based on the similarity of coverage pattern was proposed to reduce test execution and auditing cost. Sam path et al. used concept analysis to cluster user sessions aimed at selecting a set of representative user sessions that covered all the base requests [15]. In their recent study [14] they proposed a prioritization of user-session based test cases by using the criteria of session lengths, frequency appearance of request sequences, and coverage of parameter-values and their interactions. They suggested that the latter two criteria should be covered as early as possible. Alshahwan and Harman presented an algorithm to repair obsolete user sessions for regression testing of web applications [1].

## 3. Problem Formulation and Approach

The aim of our methodology is to create an adequate test suite from the collected user sessions. A test suite is considered adequate if
 (1) Every service provided by the application is executed by at least one test case;
 (2) Every fault-sensitive path is exercised by at least one test case.
 A user session is a sequence of requests made by a single user using a single client to access a web server, where a request consists of a URL *and variable name –value pairs*. User sessions are recorded during the production phase of web applications. Depending on the time period of the logging, the collected user sessions can be as large as a few gigabits of data, or a small number of similar sessions that share same URLs and variables, but different values, or they can become obsolete to the current system due to changes of services. Thus, to utilize user-session data in testing, it is necessary to systematically manage existing and incoming user sessions. Clustering user-session data is the prevalent way to partition the user sessions based on certain predefined commonality among the sessions. Elbaum et al. [9] group user sessions by the similarity of coverage patterns and apply a hierarchical agglomerative approach and Euclidean distance to measure the degree of the similarity. Sam path et al. [15] apply a concept analysis to cluster user sessions that have a common pattern of base requests. Both approaches segregate user sessions based on the patterns of the usage, which are strongly influenced by the users' behaviour, and may not reflect the actual designated system behaviour. We partition the user sessions from the system's perspective, where a system's service profile is used to cluster the user sessions to ensure the coverage of the provided services. The user sessions in the same service cluster have the most common web pages (URL_names) with the service, but they may have different sequences of executions. The flow-insensitive clustering approach is to avoid having a potentially large number of clusters. Nevertheless, some sequences of execution may be more fault-prone than the others, if the pages in the user sessions have data dependence relationships. Therefore, the selection of these critical user sessions from each cluster is crucial. The next core step of our methodology is to select a smallest set of user sessions from each cluster that covers all the dependence relationships and web pages of the user sessions in the cluster. We first construct a dependence graph depicting the data and link dependence relationships between the web pages in the application. By traversing the graph, we can compute the number of dependent pages for each page. Based on this information, the target set of the user sessions can be selected from each cluster. The selected user sessions may not include all the dependence relationships and pages in the application.

The final step to construct an adequate test suite is to augment the selected user sessions with additional requests not included in the original user sessions but carrying crucial information. Furthermore, if a service is not exercised by any collected user session; additional test cases can be created manually or by using our automatic test case generation technique presented in [6]. This can be considered a very rare case, because after adding all the reachable nodes in the graph, most likely all the web pages in all the services will be included in the test cases. This paper focuses on clustering and tailoring user sessions for test case generation. The discussion of other automatic or manual testing techniques is beyond the scope of this paper.

## 4. Design of the Proposed System.

### 4.1. User Session Clustering

A service profile describes the services provided by the application. Each service is denoted by a collection of URLs and their associated variable names (URL_names). A service profile can be obtained from the functional requirement specifications for industrial applications. Software functional specifications describe the behaviour of the system; they are crucial information for developing reliable software that meets user expectations. Specifications can be written in many different ways at various levels of formalism.

The semi-formal use case modeling has been broadly adopted in a wide range of application domains, especially by web application developers. Our approach applies use case models to obtain service profiles; other forms of specifications can also be adopted with some minor processing steps.

A use case is a description of a set of sequences of events, including variants that a system performs to yield an observable result of value to an actor. A scenario is an instance of a use case, which depicts a sequence of steps describing the interactions between the actors and the system. For web applications, each step in the scenarios normally contains the information of URLs and name variables associated with the event. Hence we can easily construct a service profile of the application from its use case model. There are systems whose specifications may not be available, such as open source software or legacy systems. To obtain a service profile of such systems, we use a reverse engineering approach to extract the information from their source code. A number of studies for reverse engineering source code to construct UML models have been proposed [3] [12] [13].

For web applications, Ricca and Tonella developed the ReWeb and Test Web tools for testing and analysis of web applications, where ReWeb downloads web pages and builds UML models, and Test Web generates test sequences and executes the test cases [16]. Di Lucca et al. also used a reverse engineering approach to construct use case models from the source code of the web applications [7], and they developed a software tool, WARE, to support this approach [8]. In addition, there are a number of web crawler software tools that apply a similar approach.

By using a service profile to cluster user sessions we can
(1) Determine how the operational usage conforms to the expected profile;
(2) Abnormal visits can be detected and analyzed for security inspection when a user session cannot match with any of the predefined service scenarios;
(3) Identify invalid user sessions after a service is changed.

To associate a user session with a service, we first define the intersection of a user session $ui$ $(1 \leq i \leq n)$ and a service $sj$ $(1 \leq j \leq m)$, $ui \cap sj$, as a set of all *URLs and variable names* which are members of both the user session $ui$ and the service $sj$. Given a user session $ui$ $(1 \leq i \leq n)$ and a set of services $S = \{s1, s2,...sm\}$, the user session $ui$ is associated with the service $sj$ $(1 \leq j \leq m)$ if $| ui \cap sj | = MAX(| ui \cap sk |)$ $1 \leq k \leq m$. In other words, a user session is associated with a service if the two have the highest number of common URLs and variable names.

If there is more than one service that satisfies $| u_i \cap s_j | =$ MAX $(| u_i \cap s_k |)$ $1 \leq k \leq m$, then one of these satisfied services would be randomly selected and the user session $u_i$ is associated with the selected service. A cluster is a collection of user sessions that are associated with the same service.

In the following, we use an example to demonstrate our methodology. The example software is an online bookstore obtained from gotocode.com, which was used in our empirical study presented in Section 4 and other studies in [9] [14].

We used a reverse engineering approach and identified eight services: *s1*: Registration,*s2*: Search, *s3*: Advance Search, *s4*: Order books, *s5*:Sign in/out, *s6*: Edit Shopping cart, *s7*: Edit user profile, and *s8*: Administration. We also collected 109 user sessions.

We used three services, *s1*, *s2*, and *s4*, and 10 user sessions (*u1*… *u10*) to demonstrate their associations. For the purpose of simplicity, only the URLs are shown in the example.

*s1*: {index.jsp ➔ Registration.jsp}
*s2*: {index.jsp ➔ Books.jsp ➔BookDetail.jsp}
*s4*: {index.jsp ➔ Login.jsp ➔ ShoppingCart.jsp ➔ShoppingCartRecord.jsp}
*u1*: {index.jsp ➔ Books.jsp ➔ BookDetail.jsp}
*u2*: {index.jsp ➔Books.jsp}
*u3*: {index.jsp ➔Books.jsp}
*u4*: {index.jsp ➔ Registration.jsp ➔ Registration.jsp ➔index.jsp}
*u5*: {index.jsp ➔ login.jsp ➔ ShoppingCart.jsp ➔ index.jsp ➔ Books.jsp ➔ BookDetail.jsp ➔ ShoppingCart.jsp ➔ShoppingCartRecord.jsp}
*u6*: {index.jsp ➔ Books.jsp➔ BookDetail.jsp ➔ShoppingCart.jsp}
*u7*: {index.jsp ➔ login.jsp➔ ShoppingCart.jsp ➔ShoppingCartRecord.jsp ➔ShoppingCart.jsp}
*u8*: {index.jsp ➔Books.jsp➔ BookDetail.jsp}
*u9*: {index.jsp ➔ Books.jsp ➔ index.jsp ➔ Books.jsp➔ BookDetail.jsp}
*u10*: {index.jsp ➔ Books.jsp➔ index.jsp ➔ Books.jsp➔ index.jsp ➔Books.jsp➔ BookDetail.jsp}
*u1* ∩ *s2* = {index.jsp, books.jsp, bookDetail.jsp}.
Therefore, $| u1 \cap s2| = 3$
*u1* ∩ *s1*= {index.jsp}. Therefore, $| u1 \cap s1| = 1$
*u1* ∩ *s4*= {index.jsp}. Therefore, $| u1 \cap s4| = 1$
Thus, *u1* is associated with *s2*. Following the same procedure, *u2*, *u3*, *u6*, *u8*, *u9*, *u10* are associated with *s1*. *u4* is associated with *s2*. *u5* and *u7* are associated with *s4*.

### 4.2 User-Session Selection and Augmentation

The next step towards user-session based test case generation is to select a representative set of user sessions from each cluster. Because our ultimate goal is not only to cover all the expected operational scenarios, but also to exercise all the dependence relationships among the web pages, our selection criterion is set on the number of dependent pages of each user session. A dependent page of a web page in a user session is the one that has a dependence relationship with the web page via link or data dependence.

A web page *A* is link dependent on a web page *B* if an execution of *B* will navigate to *A*. For example, in an index.html page, when a user clicks on the contact page link, the user will be directed to the contact page, so the contact page is link dependent on the index page. A web page *A* is data dependent on a web page *B* if the value of a variable referenced in *A* is defined in *B*, and a definition clear path exists between the two pages.

For example, in a typical e-business application, a web page, *update.html*, is used to update the shopping cart and another web page, *view.html*, is used to show the items in the shopping cart.*View.html* is data dependent on *update.html*, because the items shown in *view.html* were defined in *update.html*. To compute the number of dependent pages for each user session, we construct a system dependence graph for the application, where a node in the graph represents a web page, and an edge in the graph represents either a data or a link dependence relationship between two pages. We have developed a software tool, WebMTA [19], which constructs system dependence graphs for web applications.

To select user sessions from the clusters, we first calculate dependence counts, which are the numbers of link and data dependent pages, for each user sessions. The dependence count for a node is the number of out edges of the node, excluding recursive edges. The dependence count for a user session is the total number of dependence counts for all the URL_names in the session. Next we sort the user sessions in each cluster in a descending order based on the data dependence count first; a second sorting based on the link dependence counts is applied when the user sessions have the same number of the data dependence counts.

Finally we select the user sessions starting from the highest order until all the URL_names in the cluster are included in the selected user sessions. For each selected user session, the corresponding node of each URL_name is marked. If the node has one or more outgoing data dependence edges, then the URL_names of the dependent nodes will be placed after the URL_name of the node, if it is not already there. This ensures that all the data dependence relationships will be covered by the test suite. After the selection process is completed, for each unmarked node, *n1*, in the graph, we select one of the marked nodes, *n2,* which can reach *n1* via a link dependence edge. The URL_name of *n1* is then inserted into the user session where the URL_name of *n2* belongs and at the location after the URL_name of *n2*. Because this page was not visited by any collected user sessions, the input values, if any, of the page will be created by using random values. Figure 1 shows a partial dependence graph of the bookstore. The dependence counts of each page are listed in Table 1 and the dependence counts for each user session are listed in Table 2. Where LDC denotes link dependence count and DDC denotes data dependence count. From the clustering phrase, we know that *s1* cluster contains *u1, u2, u3, u6, u8, u9,* and *u10. s2* cluster has *u4,* and *s4* cluster includes *u5* and *u7* .

Based on the total LDC and DDC for these user sessions (*u1… u10*), the user sessions in these clusters are sorted in the following order: *s1* cluster: (*u6, u8, u9, u10,u1, u2, u3* ), *s2* cluster: (*u4)*, and *s4* cluster: (*u5 , u7* ).During the selection phrase, in *s1* cluster, *u6* was selected first, and because *u6* already covers all the URL_names in *s1* cluster, no more session was needed.*s2* cluster only has one user session *u4,* so *u4* was selected. For *s4* cluster, we selected *u5* first, and no more session was selected, because *u5* covers all the URL_names in *s4* cluster.

Therefore, *u4, u5, u6* were used to construct test cases. In the augmentation phase, we need to append admin.jsp, faq.html, MyInfo.jsp to one of the user sessions *u4, u5, u6* because they are reachable from the node (index.jsp) in sessions *u4, u5, u6.*

## 5.Implementation

An open source online bookstore obtained from gotocode.com. The online bookstore allows users to register, search, browse, modify shopping carts and purchase books online. It also allows the administrator to manage books, information of members, credit cards etc. We adopted the JSP implementation of an online bookstore, which contains 29 JSP files and 8 database tables, and added a faq.html page to assist visitors in using the system. Tomcat 6.0 was used to host the application, in which the configuration was modified to log the request data. (e.g., IP, time stamp, http request).

Firstly I took 10 user sessions to check the methodology and implemented it and the details about the dependence graph and its each page count is shown in the tables according to the access by users.

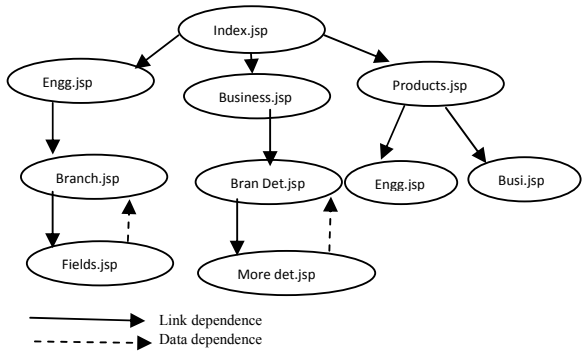

**Fig 1: Dependence Graph of Bookstore**
The figure 1 depicts neatly the dependence relationship between the pages in the Bookstore Application. By means of applying the methodology given in the section 3 and also considering the services and the user sessions the counts for each page and the user sessions are denoted from the tables given below which has been generated from the online application that has been developed.

| NODES | DDC | LDC |
|---|---|---|
| Index.jsp | 0 | 3 |
| Product.jsp | 0 | 2 |
| Engg.jsp | 0 | 1 |
| Business.jsp | 0 | 1 |
| Business details.jsp | 0 | 1 |
| Fields.jsp | 1 | 0 |
| More details.jsp | 1 | 0 |
| Branches.jsp | 0 | 1 |

**Table 1: Dependence counts of each page**

The corresponding user session's counts are calculated by the no of times a user is accessing a page that is a URL is as follows:

| User session | DDC | LDC |
|---|---|---|
| User1 | 0 | 5 |
| User2 | 1 | 5 |
| User3 | 0 | 5 |
| User4 | 0 | 5 |
| User5 | 0 | 7 |
| User6 | 1 | 7 |
| User7 | 0 | 7 |
| User8 | 0 | 7 |
| User9 | 1 | 9 |

**Table 2: Dependence counts of the user session**

By means of applying the previous methodologies that has been developed the system reduces considerable numbers of test suite only but from this implementation it has been proved that apart from all systems and methods our system reduces 15%more test cases and taking only 10 user sessions with URL_Request we got a result that according to method of concept analysis User sessions 2 and 6 are taken in order to test the application while in our method it has been reduced to one session that is it's enough to test only the user session 9 by means of clustering the user session and it has been clearly depicted in the performance analysis graph Table 3:which is result of our implementation likewise by adding more and more user sessions we will yield the better result from our experiment.

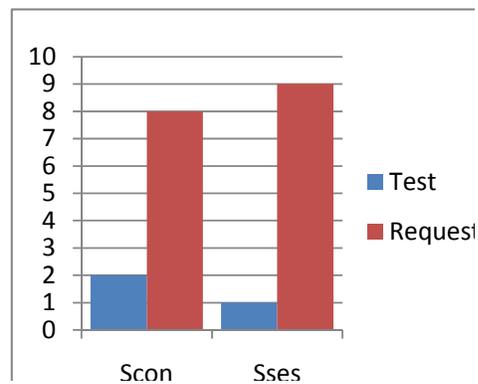

**Table 3: Performance Analysis**

## 6. Conclusion

User-session based web testing has become attractive to industries due to its use of field data, which not only reflects actual usages of the application, but also reduces the effort for test case generation. A number of studies have been conducted to demonstrate various approaches to apply this technique. The majority of these studies focus on the clustering and reduction of user-session data. Although there have been some attempts to use structural coverage criteria, including statements/blocks, conditions, and functions to address the adequacy issue, the results either are inconclusive or do not show any significant improvement of the fault detection rates [9].

In this paper, we present an approach that uses applications' service profiles to cluster user sessions, selects the user sessions from each cluster that have the largest data and link dependent pages and include all the web pages in the cluster, and then augments the selected user sessions by appending the requests to these dependent pages. The use of service profiles in clustering user sessions can facilitate the determination of how much usage has covered the expected service scenarios. The test suite selected from each cluster can best represent the requirements of the application. Moreover, by appending the requests that have dependence relationships with the requests in the user sessions, it enables the coverage of the structure of the implementation.

The results of our controlled empirical study show that all the reduction techniques performed well, as the size of the user sessions was dramatically reduced while the fault detection rates remained the same as that of the original user sessions. However, the fault detection capability of the original user sessions varies when the distributions of the faults are different. This observation implies that test cases created from the user sessions cannot guarantee a good fault detection rate. But with the augmentation, the user sessions can perform much better than the original and are independent of the fault distributions. In the second study, which is the first study conducted on an industrial system, we observed very interesting results.

The existing reduction techniques only reduced a small portion of the requests from the original user sessions. Our clustering approach was able to reduce the number of requests and tests to 15% and achieve the same fault detection rate as the original user sessions. The test suite created with the augmentation detected 99% of the faults in the system, demonstrating a great potential for effective testing of industrial web applications.

## 7. Future Work

The contributions of this paper not only present a new, effective user-session based testing technique. More importantly, our discovery of the differences between the results of the control experiment and those of the industrial system using the same existing techniques sets

i) A new milestone for future research directions and industrial practice in user-session based testing.
ii) To detect more and more faults
iii) In the future, we plan to investigate issues in regression user-session-based testing of Web applications, such as identifying and removing obsolete test cases and test case prioritization.

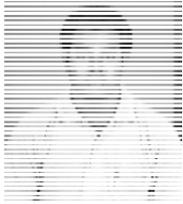
Mr.T.Deenadayalan received his B.E. degree in Computer science Engg in 2005 from Anna University Chennai and M.E. degree in Computer Science and Engineering in 2010 from Anna University Tirunelveli, Tirunelveli, India. His areas of interest are Software Testing, Digital Image Processing, Computer Networks and Soft Computing. He has presented many papers in National and International Conferences in various fields. As part of this paper, he is working on developing a tool for a specification based testing by means of applying formal methods and to reduce test suite by applying concept analysis. He is a member of ISTE.

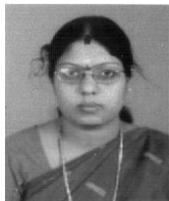
Dr.V.Kavitha obtained her B.E degree in Computer Science and Engg in 1996 from MS University and ME degree in Computer Science and Engineering in 2000 from Madurai Kama Raj University. She is the University Rank Holder in UG and Gold Medalist in PG.She received PhD degree in computer science and Engg from Anna University Chennai in 2009. Right from 1996 she is in the Department of Computer Science & Engg under various designations. Presently she is working as Asst. Prof in the Department of CSE at Anna University Tirunelveli. In addition she is the Director In-Charge of University V.O.C College of Engineering, Tuticorin. Currently, under her guidance ten Research Scholars are pursuing PhD as full time and part time. Her research interests are Wireless networks Mobile Computing, Network Security, Wireless Sensor Networks, Image Processing, Cloud Computing .She has published many papers in national and International journal in areas such as Network security, Mobile Computing, wireless network security, and Cloud Computing. She is a life time member of ISTE.

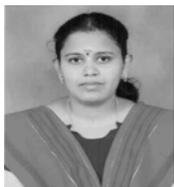
Mrs. S.Rajarajeswari received her B.E degree in Computer Science and Engg in 2003 from Madurai Kama Raj University and ME degree in Computer Science and Engineering in 2005 from Anna University Chennai. Currently she is Pursuing her PhD from Anna University Tirunelveli.She has published many papers in various fields. Her research area includes Network Security, Mobile Computing and Computer Networks. She is a member of ISTE.